\documentclass[aps,pre,preprint,floatfix]{revtex4-1}

\usepackage{amsmath}
\usepackage{amssymb}
\usepackage{graphicx}
\usepackage{dcolumn}
\usepackage{bm}
\usepackage[dvipsnames]{xcolor}
\usepackage{multirow}
\usepackage{hyperref}
\usepackage{epstopdf}

\usepackage[utf8]{inputenc}
\usepackage[T1]{fontenc}
\usepackage{microtype}
\usepackage{txfonts}
\usepackage{ulem}

\begin{document}


\title{Charge regulation radically modifies electrostatics in membrane stacks}

\author{Arghya Majee$^{1,2}$}
\email{majee@is.mpg.de}

\author{Markus Bier$^{1,2,3}$}

\author{Ralf Blossey$^4$}

\author{Rudolf Podgornik$^{5,6}$}
             
\affiliation
{
$^{1}$Max-Planck-Institut f\"ur Intelligente Systeme, Heisenbergstr.\ 3, 70569 Stuttgart, Germany\\ 
$^{2}$IV. Institut f\"{u}r Theoretische Physik, Universit\"{a}t Stuttgart, Pfaffenwaldring 57, 
      70569 Stuttgart, Germany\\
$^{3}$Fakult\"at Angewandte Natur- und Geisteswissenschaften, 
      Hochschule f\"ur Angewandte Wissenschaften W\"urzburg-Schweinfurt, 
      Ignaz-Sch\"on-Str.\ 11, 97421 Schweinfurt, Germany\\
$^{4}$Universit\'e de Lille, CNRS, UMR8576 Unit\'e de Glycobiologie Structurale et Fonctionnelle (UGSF), 
      F-59000 Lille, France\\
$^{5}$School of Physical Sciences and Kavli Institute for Theoretical Sciences, 
      University of Chinese Academy of Sciences, Beijing 100049, China\\
$^{6}$CAS Key Laboratory of Soft Matter Physics, Institute of Physics, Chinese Academy of Sciences, 
      Beijing 100190, China\\ 
}

\date{November 6, 2019}

\begin{abstract}
Motivated by biological membrane-containing organelles in plants 
and photosynthetic bacteria, we study charge regulation in a model 
membrane stack. Considering (de)protonation as the simplest mechanism 
of charge equilibration between the membranes and with the bathing 
environment, we uncover a symmetry-broken charge state in the stack 
with a quasiperiodic effective charge sequence. In the case of a 
monovalent bathing salt solution our model predicts complex, 
inhomogeneous charge equilibria depending on the strength of 
the (de)protonation reaction, salt concentration, intermembrane 
separation, and their number in the stack. Our results shed light 
on the basic reorganization mechanism of thylakoid membrane stacks. 
\end{abstract}

\maketitle


\section{Introduction}

The charge regulation mechanism, described originally in 
the 1920's \cite{Lin24} and later developed by Kirkwood 
and Schumaker \cite{Kir52}, Marcus \cite{Mar55} and Lifson 
\cite{Lif57} has become a topic of considerable research 
interest in recent years \cite{Lun13, Adz15, Pod18, Avn18, San19}. 
Charge regulation refers to the situation in which the 
local charge on a solvated surface responds to changes 
in the environment, such as local pH, dielectric inhomogeneities, 
salt concentrations, etc. The presence of dissociable 
groups, in particular on biological surfaces, then allows 
the surface charge to adapt to local and global solution 
conditions. As a consequence, soft-matter electrostatics, 
often formulated within the Poisson-Boltzmann (PB) paradigm, 
needs to be implemented with the self-consistent boundary 
condition \cite{Nin71}, superseding the assumptions of 
constant charge vs constant potential dichotomy 
\cite{Mar16}. However, recent work uncovered charge 
regulation phenomena that cannot be rationalized even within 
the modified boundary conditions framework \cite{Maj18}. In 
fact, the interaction of two planar, chargeable surfaces in 
a bathing electrolyte was shown to be much richer than what 
could be predicted based on the constant charge vs constant 
potential phenomenology: the charge symmetry itself can become 
broken and the interaction turning attractive instead of being, 
as expected otherwise, repulsive. This also ties in with recent 
proposals that the attractive non-DLVO (Derjaguin-Landau-Verwey-Overbeek) 
forces between colloidal surfaces, exhibiting adsorption 
and dissociation equilibria are unrelated to ion–ion 
correlations but depend rather directly on the charge 
regulation mechanisms \cite{Bor18}.

Reconstructed lamellae and bilayer systems with fixed charges 
are well understood both experimentally and theoretically 
\cite{Mar19, Her14, His17}. Biological membrane stacks, however, 
generally have a much more complex electrostatics. We take 
thylakoid membranes in plants and photosynthetic bacteria as 
examples. They typically consist of $\approx 10$ bilayers, stacked 
on top of each other to form grana, the light-harvesting 
organelles. The organization of these lamellae is obviously 
highly heterogeneous, since they are the carriers of the 
photosynthetic proteins. Thylakoid stacking has been shown long 
ago to be driven by electrostatics \cite{Bar80-1, Rub80, Bar82}, 
and up to only recently, the classic PB paradigm has been 
implemented and extended to quantify experiments on thylakoid 
stacking \cite{Put17}. However, the charges at the thylakoid 
membranes are obviously not constant, but depend on the 
equilibration through two relevant processes: (de)protonation 
and (de)phosphorylation \cite{All82, Kan16}, frequently 
leading to the observation of asymmetrically charged 
membranes \cite{All82, Kan16}. Given that the buildup 
of grana by thylakoids is dynamic and light dependent, 
and very little is still known about the regulation of this 
process \cite{Mus08, Kir11, Kir18}, a better understanding 
of the effect of charge regulation on membrane stacks is 
generally called for. Physical theory can play a crucial role, 
as we show by elucidating the effect played by charge regulation 
in a membrane stack, in which the embedding environment is 
taken into account. 

\begin{figure}[!t]
\begin{center}
\includegraphics[width=0.5\columnwidth]{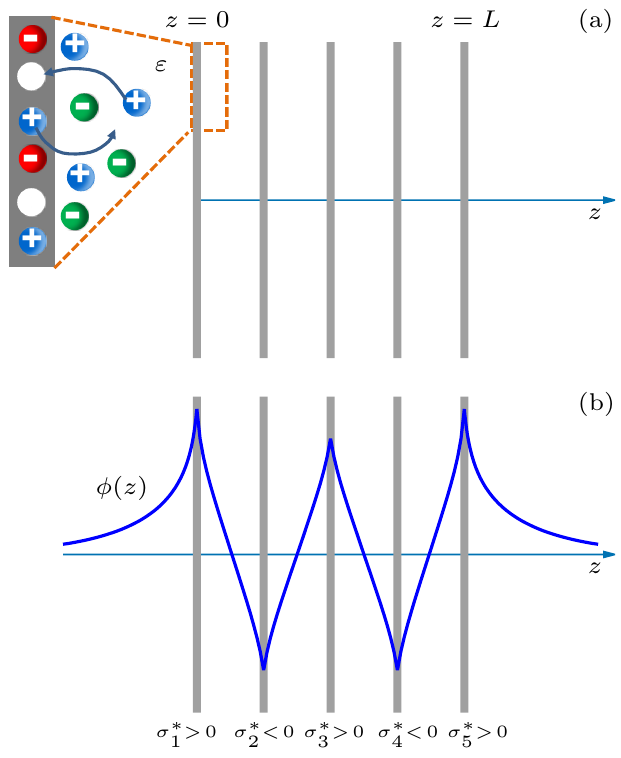}
\caption{(a) Sketch of our system consisting of an array of $N$ 
             charge-regulated, equally-spaced, parallel surfaces immersed 
             in a solvent of permittivity $\varepsilon$. The case depicted 
             here corresponds to $N=5$. The surfaces are placed
	     perpendicularly to the $z$-axis with the first and the last
	     ones being situated at $z=0$ and $z=L$, respectively. As shown
	     under magnification, each surface contains fixed negative charges
	    (red circles) and neutral sites (white circles). The surfaces are
	     charge regulated via (de)protonation (blue circles) of these 
	     sites. The green circles indicate the mobile anions in the fluid. 
	    (b) Under certain circumstances the surfaces can become 
	     alternatingly positively ($\sigma_{\!j}^{*}>0$) and negatively 
	    ($\sigma_{\!j}^{*}<0$) charged; the corresponding electrostatic 
	     potential profile $\phi(z)$ is shown by the blue line.}
\label{Fig1}
\end{center}
\end{figure}

\section{Model and formalism}

We consider an array of $N$ charge-regulated, equally-spaced, parallel 
membranes of negligible thickness \cite{finite} immersed in a solvent of permittivity 
$\varepsilon\,(=\varepsilon_r\varepsilon_0$ with the relative permittivity 
$\varepsilon_r$ and the permittivity of vacuum $\varepsilon_0)$ as depicted 
in Fig.~\ref{Fig1}. In a three-dimensional Cartesian coordinate system 
the first surface is located at $z=0$ and the $N$-th surface at $z=L$, 
implying a separation $\Delta L=L/\left(N-1\right)$ between consecutive 
surfaces. As it is the case for proteins or membranes, the surfaces 
consist of fixed negative charges with surface number density $M$ and 
neutral sites with surface number density $\theta M=1/a^2$ where 
(de)protonation can occur. Here we consider the case $\theta=2$ for which 
the surface charge density in units of $e/a^2$ on the $j$th surface can 
vary in the interval $\sigma_{\!j}^*\in\left[-\frac{1}{2},\frac{1}{2}\right]$ 
(see \cite{Maj18}). 

The grand potential per unit surface area in the units of $k_BT=1/\beta$ 
corresponding to our system is then given by \cite{Maj18} (see also \cite{Har06, Tag10})
\begin{align}
& \beta\Omega\left[\sigma^{*}\right]=-\frac{\varepsilon}{\beta e^2}\int\limits_{-\infty}^{\infty}dz
 \left[\kappa^2\left(\cosh\left(\phi(z)\right)-1\right)+\frac{1}{2}\left(\phi'(z)\right)^2\right]\label{eq:1}\\
&\!\! \!\!+ \frac{1}{a^2}\sum\limits_{j=1}^{N}\Bigg[\sigma_{\!j}^*\phi_j
 -\alpha\eta_j-\frac{\chi \eta_j^2}{2}
 +\eta_j\ln\eta_j+\left(1-\eta_j\right)\ln\left(1-\eta_j\right)\Bigg],\nonumber
\end{align}
where $\phi$ is the dimensionless electrostatic potential satisfying 
the PB equation $\phi''=\kappa^2\sinh\phi$ (see \cite{Rus89}) 
with the inverse Debye length $\kappa$ and the Neumann boundary 
conditions at the surfaces set by $\sigma^*=(\sigma_{\!1}^*,\dots,\sigma_{\!N}^*)$. 
The electrostatic potential and the fraction of sites occupied by 
protons at the $j$th surface are given by $\phi_j$ and $\eta_j$, 
respectively, where the latter obeys the relation $\eta_j=\sigma_{\!j}^{*}+1/2$. 
The terms containing $\phi$ in Eq.~\eqref{eq:1} correspond to the 
electrostatic field energy and the logarithmic terms are an entropic 
contribution. As in Ref.~\cite{Maj18}, the term $-\alpha\eta_j$ 
describes the non-electrostatic adsorption free energy penalty per 
ion and the term $-\frac{1}{2}\chi\eta_j^2$ proportional to the 
Flory-Huggins parameter $\chi$ is the change in the nonelectrostatic
interaction between adsorbed ions on neighboring sites upon (de)protonation. 
Nonelectrostatic interaction between adsorbed ions may arise 
from the formation of complex hydrogen bonded network of water 
molecules or from forces of quantum-chemical origin such as van 
der Waals (vdW) and/or chemical bonding interactions, that could be 
of repulsive as well as attractive nature. The equilibrium values 
of $\sigma_{\!j}^*$ minimize the grand potential
$\beta\Omega\left[\sigma^{*}\right]$ in Eq.~\eqref{eq:1} (see 
Appendix for details).

\begin{figure}[!t]
\includegraphics[width=0.5\columnwidth]{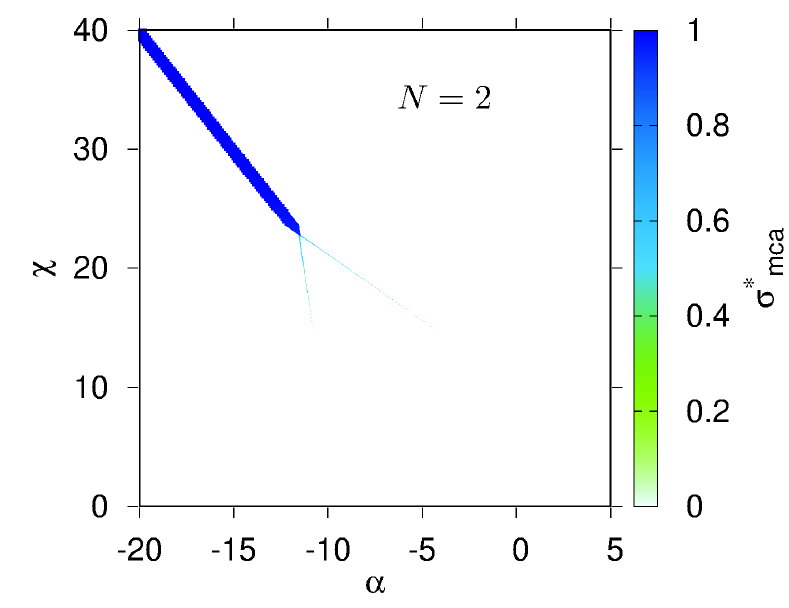}
\caption{Variation of the quantity $\sigma_{\text{mca}}^*$ 
        (defined in Eq.~\eqref{eq:2}) with the parameters $\alpha$ 
        and $\chi$ for $N=2$ charge-regulated interacting 
        surfaces immersed in a solvent. The surfaces are considered 
        to be fixed in space with a dimensionless separation 
        $\kappa\Delta L\approx1$.}
\label{Fig2}
\end{figure}

\section{Results and discussion}
The mean charge asymmetry
\begin{align}
 \sigma_{\text{mca}}^*=\frac{1}{N-1}\sum\limits_{j=1}^{N-1}\left|\sigma_{\!j}^*-\sigma_{\!j+1}^*\right|\in[0,1]
 \label{eq:2}
\end{align}
allows one to distinguish between equally
$(\sigma_{\text{mca}}^*=0)$ and unequally $(\sigma_{\text{mca}}^*>0)$ charged 
neighboring surfaces of the stack. The maximal value $\sigma_\text{mca}^*=1$ 
corresponds to a configuration with all surfaces being completely charged 
($|\sigma_j^*|=\frac{1}{2}$) with alternating positive and negative signs.
In the following, we discuss the behavior of $\sigma_{\text{mca}}^*$ for 
a varying number of surfaces in the stack as well as for varying strengths 
of the parameters $\alpha$ and $\chi$. Unless stated otherwise, we consider 
surfaces with $a=1\,\mathrm{nm}$ immersed in an aqueous electrolyte solution 
$(\varepsilon_r\approx80)$ at $T=300\,\mathrm{K}$. Please note that 
for these values of $\varepsilon_r$ and $T$, an ionic strength of $10\,\mathrm{mM}$ 
results in a Debye length $\kappa^{-1}\approx3\,\mathrm{nm}$.

First we consider the simplest case of $N=2$ surfaces, which differs 
from the situation considered in Ref.~\cite{Maj18}, where the regions 
outside the stack were not included. Their presence may influence the 
charge regulation at the surfaces and in turn affect the charge 
asymmetry observed in Ref.~\cite{Maj18}, which results from a tradeoff
among the energy costs due to ion-surface, in-plane ion-ion and
surface-surface interactions. However, as one can infer from the 
colored regions in Fig.~\ref{Fig2}, depending upon the values of 
the parameters $\alpha$ and $\chi$, the surfaces can still be 
unequally charged. For $\alpha<\alpha_0\,(\approx-11.4)$ 
a dark blue region appears around the line $\chi=-2\alpha$ with 
$\sigma_{\text{mca}}^*\approx1$, implying that the surfaces are 
oppositely charged with $|\sigma_{\!1}^*|\approx|\sigma_{\!2}^*|\approx\frac{1}{2}$. 
For $\alpha>\alpha_0$, two very thin tails with lower asymmetry 
between $\sigma_{\!1}^*$ and $\sigma_{\!2}^*$ appear on either 
side of the line $\chi=-2\alpha$. Qualitatively, these results 
are in perfect agreement with Fig.~2(b) of Ref.~\cite{Maj18}.
The outside regions influence the asymmetric charge regulation only 
quantitatively, e.g., by giving rise to a thinning of the two tails 
in Fig.~\ref{Fig2} as compared to Fig.~2(b) of Ref.~\cite{Maj18}. 

\begin{figure}[!t]
\includegraphics[width=0.5\columnwidth]{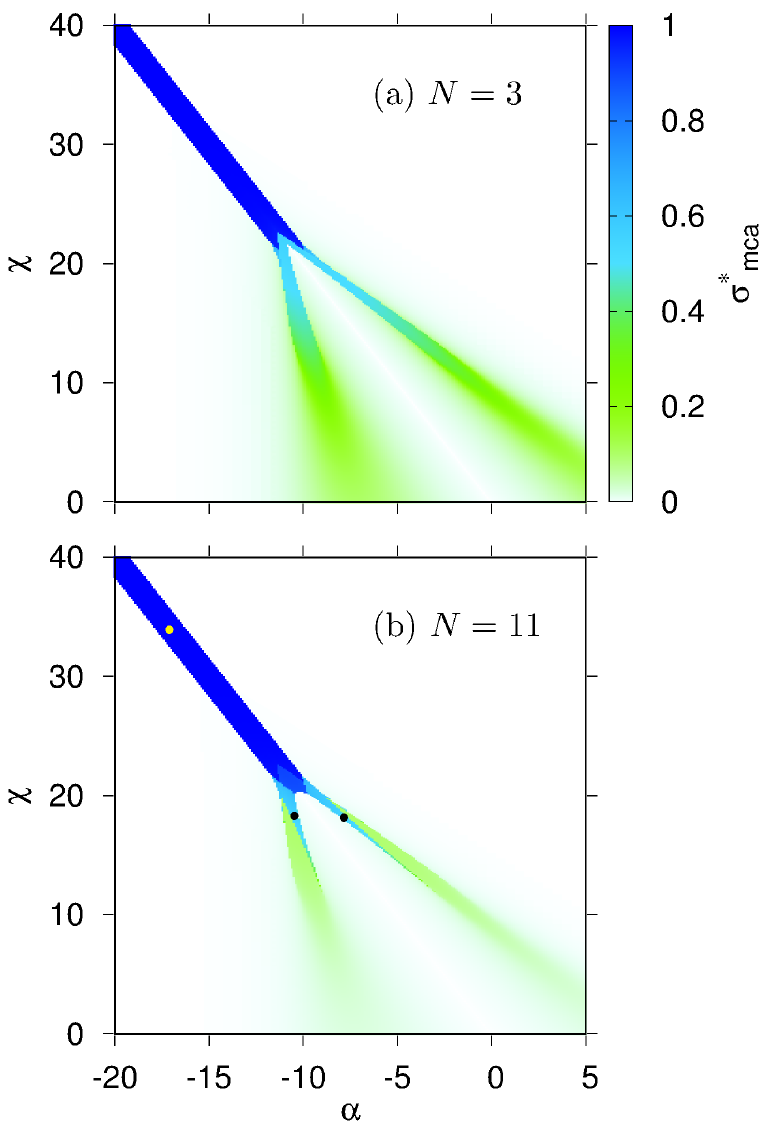}
\caption{Variation of the quantity $\sigma_{\text{mca}}^*$ (defined 
         in Eq.~\eqref{eq:2}) with the interaction parameters $\alpha$ 
         and $\chi$ for stacks with (a) $N=3$ and (b) $N=11$ charge-regulated 
         interacting surfaces immersed in a solvent. The surfaces in 
         each case are considered to be fixed in space with a dimensionless 
         separation $\kappa\Delta L\approx1$ between two consecutive 
         surfaces. The yellow (light) and the black bullets in the lower panel 
         represent points considered in Figs.~\ref{Fig5} and \ref{Fig6}.}
\label{Fig3}
\end{figure}

\begin{figure}[!t]
\includegraphics[width=0.5\columnwidth]{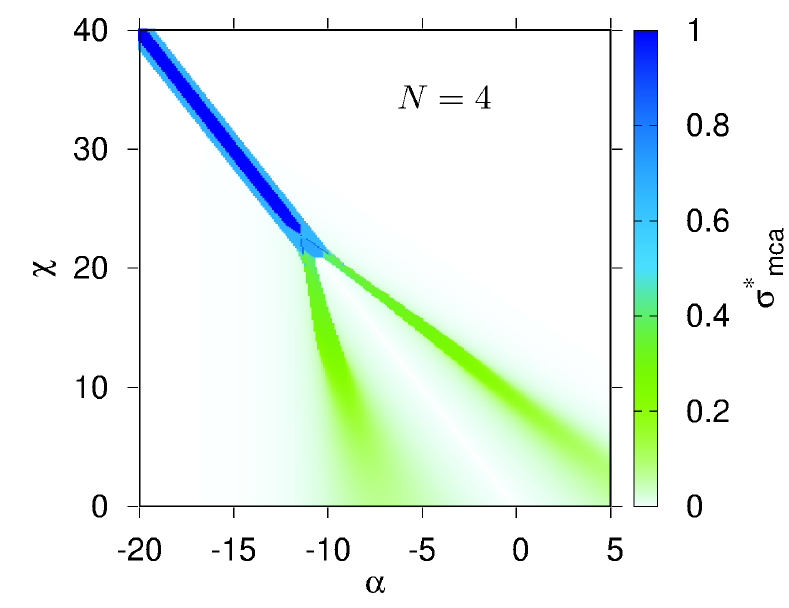}
\caption{Variation of the quantity $\sigma_{\text{mca}}^*$ (defined 
         in Eq.~\eqref{eq:2}) with the interaction parameters $\alpha$ 
         and $\chi$ for a stack of $N=4$ charge-regulated interacting 
         surfaces immersed in a solvent. The surfaces are
         considered to be fixed in space with a dimensionless separation 
         $\kappa\Delta L\approx1$ between two consecutive surfaces.}
\label{Fig4}
\end{figure}

Now we consider stacks with more than two surfaces. As pairwise 
electrostatic attractions between consecutive surfaces promote a 
lowering of the grand potential of the system, one would naturally 
expect the occurrence of alternatingly positively and negatively
charged surfaces within the stack, similar to the case of two 
surfaces. Consequently, the local electrostatic potential 
would exhibit ``oscillations'' (see Fig.~\ref{Fig1}(b) for an 
example with $N=5$ surfaces) that would modify the value of the 
Donnan potential \cite{Bas93}, making it effectively vanish.  
Compared to the case of  $N=2$ in Fig.~\ref{Fig2}, the colored 
regions in the $\alpha$-$\chi$ plane are broadened for $N>2$ 
(see Figs.~\ref{Fig3} and \ref{Fig4}).

Figure~\ref{Fig3} shows the mean charge asymmetry $\sigma_\text{mca}^*$ 
for odd numbers $N$ of surfaces in the stack. Upon increasing $N\geq3$, 
the dark blue region with the highest asymmetry remains almost 
unaffected but the tails shrink. Inside the dark blue region close 
to the line $\chi=-2\alpha$, the entropic terms in Eq.~\eqref{eq:1} 
almost vanish and the terms involving $\alpha$ and $\chi$ nearly 
cancel each other such that the electrostatic attraction between
consecutive surfaces gives rise to the dominant contribution to the 
grand potential. The key observation is that this electrostatic
contribution is invariant upon exchanging all signs of the 
surface charges. By virtue of a slight preference for filled or 
empty sites for $\chi\gtrsim -2\alpha$ or $\chi\lesssim-2\alpha$, 
respectively, one observes crossovers between different charging 
patterns upon changing $\chi$. For example, for $N=3$ (see 
Fig.~\ref{Fig3}(a)), the charge distribution changes from $(-,+,-)$ 
with a net negative charge for $\chi<-2\alpha$ to $(+,-,+)$ with 
a net positive charge for $\chi>-2 \alpha$. Here `$+$' and `$-$' 
represent charge densities $\sigma_j^*$ of approximately $+\frac{1}{2}$ 
and $-\frac{1}{2}$, respectively. As $\alpha$ is increased beyond 
the dark blue region, two more regions (tails) with lower charge 
asymmetry are obtained. For $N=3$, in the upper (bluish) part of the 
lower tail (below the line $\chi=-2\alpha$) in Fig.~\ref{Fig3}(a), 
the charge distribution is $(0^+,-,0^+)$, where $0^+$ represents a 
slightly positive charge density. Along the lower tail, the surfaces 
with charge densities $0^+$ become more negatively charged. However, 
their charge densities differ from the middle surface within the 
entire greenish region. Exactly the same phenomenon is observed 
within the upper tail albeit with $0^+$ and $-$ being replaced 
by $0^-$ and $+$, respectively. Moreover, similar trends are 
observed for $N=11$ (see Fig.~\ref{Fig3}(b)) as well.

Next, we consider the case of even numbers $N$ of surfaces in the 
stack (see Fig.~\ref{Fig4}). A significant difference compared to 
odd $N$ is observed regarding the charge distribution among the 
surfaces within the dark blue region close to the line $\chi=-2\alpha$. 
For even $N$ and alternatingly charged surfaces the total charge 
of all the surfaces is always zero. Therefore, with increasing 
$\chi$, a mere flipping of signs of the charges on each surface is 
not favorable as it does not allow adsorption of more protons onto 
the surfaces. Consequently, a competition between the electrostatic 
interaction energy and the contribution of the term involving $\chi$ 
in Eq.~\eqref{eq:1} takes place which leads to a very different 
reorganization of charges among the surfaces. For example,
within the blue region of Fig.~\ref{Fig4}, the distribution of 
charges inside the stack of $N=4$ surfaces changes from $(-,+,-,-)$ 
for $\chi<-2\alpha$ via $(-,+,-,+)$ for $\chi\approx-2\alpha$ to 
$(+,+,-,+)$ for $\chi>-2\alpha$. Although in the configurations 
$(-,+,-,-)$ and $(+,+,-,+)$ one pair of surfaces repels each other
electrostatically, they are favorable due to the ion-ion interaction 
at each surface. Inside the lower tail in Fig.~\ref{Fig4} the charge
distribution is of the type $(0^+,-,0^+,0^-)$ or $(0^+,-,0^-,0^-)$ 
in the upper part whereas it is of the type $(0^-,-,-,0^-)$ in the 
lower part. For the upper tail, one just needs to invert the signs. 
Exactly the same phenomenon occurs for $N=10$ (see Fig.~\ref{Fig1_SI} 
in the Appendix). So far we have considered the case 
$\kappa\Delta L\approx1$ only. With decreasing separation between 
the surfaces, the electrostatic pair interaction becomes stronger 
and the dark blue region spreads to dominate over the tails (see 
Fig.~\ref{Fig1_SI} in the Appendix).

\begin{figure}[!t]
\includegraphics[width=0.5\columnwidth]{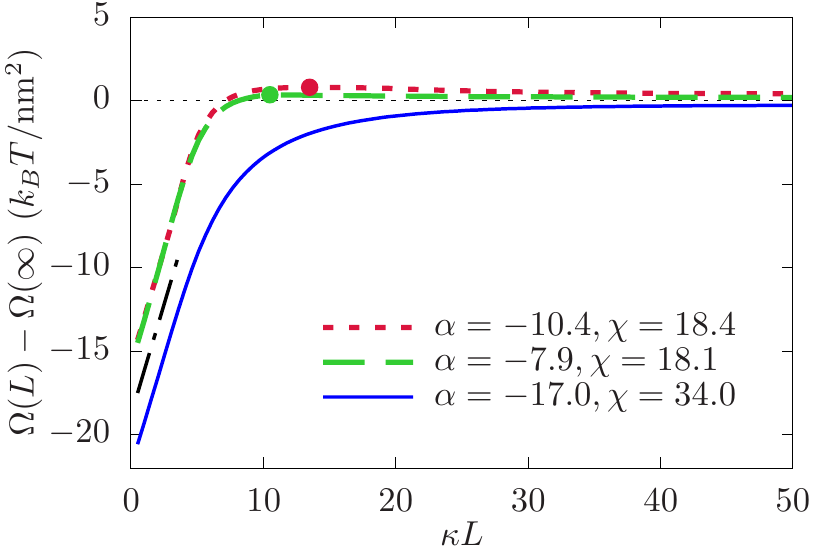}
\caption{Binding energy per cross-sectional area of a stack
         of $N=11$ surfaces as a function of the dimensionless
	 stack width $\kappa L$ for three different combinations of 
	 the interaction parameters $\alpha$ and $\chi$. The blue 
	 solid, red dotted, and green dashed lines correspond to 
	 points in the dark blue region, the lower tail, and the 
         upper tail of Fig.~\ref{Fig3}(b), respectively. The
	 binding energy increases within the entire considered 
	 range of $\kappa L$ for the blue curve as the electrostatic 
	 interaction is strongest there. For the other cases, the 
	 electrostatic interaction is weaker such that tiny 
	 barriers (marked by the large dots) occur. The dash-dotted 
	 straight line depicts a slope of $2.66\,k_BT/\mathrm{nm^2}$, 
	 which corresponds to the asymptotic disjoining pressure at 
	 small stack widths $\kappa L\lesssim4\ll N-1=10$ and 
	 $\kappa\approx0.1027\,\mathrm{nm}^{-1}$.}
\label{Fig5}
\end{figure}

\begin{figure}[!t]
\includegraphics[width=0.5\columnwidth]{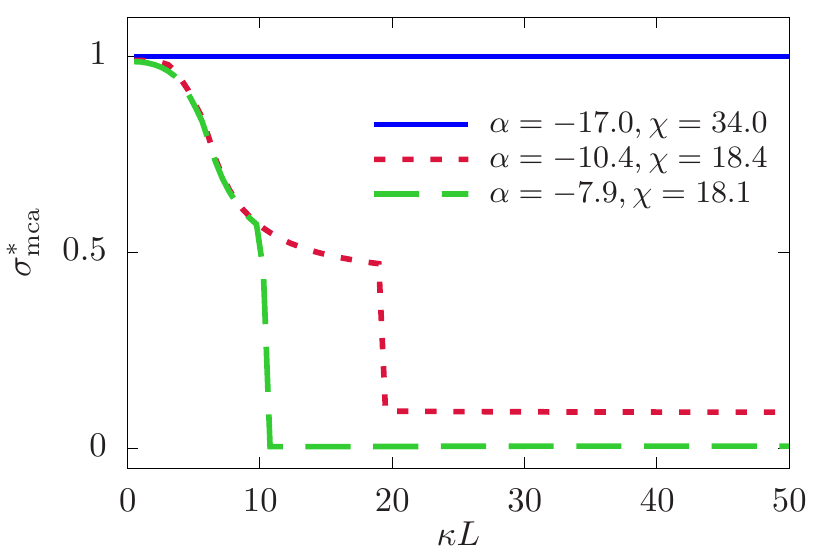}
\caption{Mean charge asymmetry $\sigma_\text{mca}^*$ for 
a stack of $N=11$ surfaces as a function of the dimensionless
	 stack width $\kappa L$ for the three combinations of 
	 the interaction parameters $\alpha$ and $\chi$ considered in 
	 Fig.~\ref{Fig5}. The blue curve corresponds to 
	 a point (marked with bullet) on the $\chi=-2\alpha$ line in 
	 Fig.~\ref{Fig3}(b) where extreme charge 
	 asymmetry leading to $\sigma_\text{mca}^*\approx1$ prevails for the 
	 entire range of separations shown here. The other two lines
	 correspond to points which, for very short separations, lie 
	 within the dark blue region, then fall within the light blue 
	 regions of the tails before making a transition to the green or 
	 white regions with likely charged surfaces.}
\label{Fig6}
\end{figure}

After calculating the equilibrium surface charge densities 
$\sigma_{\!j}^{*}$ for fixed parameters $\alpha$, $\chi$, and $N$ 
as functions of the stack width $L$ and using Eq.~\eqref{eq:1}, one 
obtains the grand potential per unit cross-sectional area $\Omega(L)$.
In the limit $L\rightarrow\infty$, the electrostatic interaction 
between the surfaces vanishes. The binding energy of the stack 
$\Omega(L)-\Omega(\infty)$ is shown in Fig.~\ref{Fig5} as a 
function of the dimensionless stack width $\kappa L$ for $N=11$ 
and three different combinations of the parameters $\alpha$ and 
$\chi$. In all three cases, the binding energy $\Omega(L)-\Omega(\infty)$ 
increases initially, implying that the disjoining pressure
$-\partial\Omega(L)/\partial L$ within the stack is attractive 
for small widths $L$. For high mean charge asymmetry 
$\sigma^*_\text{mca}\approx1$ and surface separations 
$\Delta L = L/(N-1) \ll 1/\kappa$ much smaller than the Debye 
length $1/\kappa$, no significant screening of the electric field by
the ions takes place such that the system can be viewed as a series 
of $N-1$ capacitors each charged with approximately half of the 
maximal surface charge density $|\sigma^*_j|/2\approx1/4$.
From the electric field energy of such a setup one infers the 
disjoining pressure $-\partial\Omega(L)/\partial L\simeq-0.273\,k_BT/
\mathrm{nm^3}=-11.3\,\mathrm{bar}$ for $L\ll(N-1)/\kappa$ (see 
Appendix for details). Note that this value, which corresponds 
to the slope depicted in Fig.~\ref{Fig5} in the range 
$\kappa L\lesssim4\ll N-1=10$, is independent of the number 
of surfaces $N$ and of the parameters $\alpha$ and $\chi$, as 
long as high mean charge asymmetry prevails. For the dashed 
and the dotted curves, which correspond to points within the 
two tails in Fig.~\ref{Fig3}(b), the binding energy exhibits a 
tiny barrier, i.e., the disjoining pressure becomes repulsive 
for sufficiently thick stacks. The disjoining pressure 
derived from the solid line, corresponding to parameters $\alpha$ 
and $\chi$ within the dark blue region in Fig.~\ref{Fig3}(b), 
is attractive throughout the entire range of $\kappa L$ values 
shown, as the electrostatic interaction is the strongest here. 
The corresponding changes in the charge patterns 
are confirmed by the variation of the mean charge asymmetry 
$\sigma_{\text{mca}}^*$ shown in Fig.~\ref{Fig6}. Whereas 
$\sigma_{\text{mca}}^*$ remains constant at the highest
value $(\approx1)$ throughout the entire range of $\kappa L$ 
for the blue solid curve, it decreases with increasing $\kappa L$ 
for the two points in the tails. With increasing separation 
between the surfaces, the colored regions with asymmetric 
charge distributions shrink. Consequently, these points first 
fall within the dark blue region with highest charge asymmetry, 
then in the sky blue region with lower asymmetry, before making 
a transition to the green tails indicating unequally but 
likely charged surfaces or to the white regions with equally 
charged surfaces.

Finally we discuss the relevance of our results in view of two 
systems, namely the plant thylakoid membrane and photosynthetic 
membranes of the family of cyanobacteria, in the same spirit as
has been done for the standard PB case, see e.g. \cite{Put17}. 
Both systems show stack-like structures in their light-harvesting 
organelles. For thylakoids, typically $\approx10$ disks of 
diameter in the range $300\dots600\,\mathrm{nm}$ are stacked 
together with interdisk separation $\Delta L\approx3.5\,\mathrm{nm}$ 
\cite{Mus08, Kir11, Put17}. The ionic strength of the embedding 
medium shows two different ranges for monovalent salts \cite{Put17}: 
$2\dots20\,\mathrm{mM}$ \cite{Scu80, Cru01, Kan16} and 
$100\dots200\,\mathrm{mM}$ \cite{Phi01}. Up to a salinity of 
$\approx10\,\mathrm{mM}$ the dimensionless stack width is given 
by $\kappa L\lesssim10$, for which the disjoining pressure 
is attractive for all asymmetrically charged configurations 
(see Fig.~\ref{Fig5}). For higher salinities, $\kappa L$ is 
beyond the barrier of the dashed and the dotted lines in 
Fig.~\ref{Fig5}. However, even for the highest salinity of 
$200\,\mathrm{mM}$ with $\kappa L\approx50$ the disjoining 
pressure is still attractive within the dark blue region. It 
has been verified, that the charge-regulation-induced attraction 
described above dominates quantitatively over the attractive 
van der Waals force, even for the recently suggested high value 
of the Hamaker constant $A\approx4.8\times10^{-20}\,\mathrm{J}$ 
\cite{Put17}, an order of magnitude higher than the standard 
estimates \cite{Ham06}. Therefore, when asymmetric charging 
of the stack occurs, the disjoining pressure is dominated by 
the electrostatic contribution which, at very short 
separations ($\lesssim2\,\mathrm{nm}$), encounters the repulsive 
hydration force \cite{Bar80-2}. Similar numbers are found for 
photosynthetic members of the family of cyanobacteria; typical 
separation distances lie within $10\dots120\,\mathrm{nm}$ \cite{Sti16, Mee12} 
and the concentration of monovalent ions falls in the range 
of $10\dots100\,\mathrm{mM}$ \cite{Van15}. 

\section{Conclusions}
To conclude, we have studied the electrostatic interaction 
of charge-regulating membranes forming a stack composed of 
$N$ membranes, embedded in a solution of monovalent salt.  
Depending on the system parameters, specifically the number 
of membranes $N$ in the stack, the charge regulation 
mechanism leads to a complex distribution of attraction 
and repulsion forces between membranes in the stack. When 
symmetry broken charge state occurs in the stack, the 
resulting attraction can easily dominate over the vdW force 
between the membrane surfaces.

While obtained within a simple model for the charge 
regulation mechanism, our results have an impact on models 
of biological membranes, in particular of thylakoid surfaces 
in grana, the organelles of photosynthesis in plants. The 
formation of these membrane stacks has been largely 
discussed in terms of simple electrostatics models based 
on the standard Poisson-Boltzmann equation for monovalent 
or divalent salts, in which the charges on the membrane 
surfaces have been related to the charge state of 
membrane-embedded protein complexes, notably their 
phosphorylation status, see, e.g., \cite{Put17}. Our 
results confirm the relevance of electrostatics on the 
stability of membrane stacks; however, they show that a 
simple balancing of the resulting repulsive forces 
against attractive vdW forces between two surfaces is 
generally not sufficient for an understanding of thylakoid 
stacks. The changes of the charge state on the membranes 
in the stacks affect the overall charge distribution 
in a complex manner which may also have a strong influence 
on the dynamics of thylakoid stack formation, a long-standing 
open problem in plant science \cite{Mus08,Kir18}, and for 
photosynthetic cyanobacteria \cite{Sti16}. Changes between 
the charge load on the membranes, which occur naturally 
as a function of illumination, are known to affect the 
packing of the stacks \cite{Kan16}. Hence, the formation of 
asymmetrically charged membranes is expected to play a major 
role in the formation of stacks. 


\setcounter{figure}{0}
\setcounter{equation}{0}

\makeatletter
\renewcommand{\theequation}{A\arabic{equation}}
\makeatother

\makeatletter
\renewcommand{\thefigure}{A\arabic{figure}}
\makeatother

\begin{centering}
\section*{Appendix}
\end{centering}

\begin{centering}
\subsection*{1. The grand potential}
\end{centering}

Treating the ions as point-like particles and ignoring ion-ion 
correlation within a mean-field formalism, the grand potential 
corresponding to our system in units of the thermal energy 
$\beta=1/k_BT$ is given by
\begin{align}
 \beta\Omega\left[\eta,\varrho_{\pm}\right]= 
    &\int\limits_Vd^3r\Bigg[\sum\limits_{i=\pm}\varrho_i\left(\mathbf{r}\right)
     \Bigg\{\ln\left(\frac{\varrho_i\left(\mathbf{r}\right)}{\zeta_i}\right)-1\Bigg\}
     +\frac{\beta\mathbf{D}
     \left(\mathbf{r},\left[\varrho_\pm,\eta\right]\right)^2}{2\varepsilon}\Bigg]\notag\\
    &+\frac{1}{a^2}\sum\limits_{j=1}^{N}
     \Bigg[-\alpha\eta_j-\frac{\chi \eta_j^2}{2}
     +\eta_j\ln\eta_j+\left(1-\eta_j\right)\ln\left(1-\eta_j\right)\Bigg],
\label{eq:1_SI}
\end{align}
where the term under the first summation describes the entropic 
ideal gas contribution formed by the ions with 
$\varrho_\pm\left(\mathbf{r}\right)$ and $\zeta_\pm$ denoting 
the number densities at position $\mathbf{r}\in V$ inside the 
electrolyte solution and fugacities of the $\pm$-ions, 
respectively, and $\eta=(\eta_{1},\dots,\eta_{N})$ denotes 
the fraction of sites occupied by protons on each surface. The 
remaining term under the integral describes the Coulomb 
interaction among all charges in the system in a mean-field-like 
fashion in terms of the electric displacement field $\mathbf{D}$. 
The remaining quantities have the same meaning as defined in the 
main text.

Minimization of the grand potential in Eq.~\eqref{eq:1_SI} with 
respect to the ionic density profiles $\varrho_{\pm}$ provides the 
Euler-Lagrange equation which ultimately leads to the PB equation 
$\phi''=\kappa^2\sinh\left(\phi\right)$ subjected to Neumann boundary 
conditions at the surfaces set by the prescribed ion adsorption 
profile $\eta$. Here $\phi$ is the dimensionless electrostatic 
potential expressed in units of $\beta e$. Hence the equilibrium ion 
density profiles $\varrho_{\pm}\left[\eta\right]$ and the equilibrium 
electrostatic potential $\phi\left[\eta\right]$ are functionals of 
the adsorption profile $\eta$. Inserting $\varrho_{\pm}\left[\eta\right]$ 
in Eq.~\eqref{eq:1_SI} one obtains the grand potential merely in 
terms of the adsorption profile $\eta$:
\begin{align}
\beta\Omega\left[\sigma^{*}\right]=&-\frac{\varepsilon}{\beta e^2}\int\limits_{-\infty}^{\infty}dz
 \left[\kappa^2\left(\cosh\left(\phi(z)\right)-1\right)+\frac{1}{2}\left(\phi'(z)\right)^2\right]\nonumber\\
&+\frac{1}{a^2}\sum\limits_{j=1}^{N}\Bigg[\sigma_{j}^*\phi_j
 -\alpha\eta_j-\frac{\chi \eta_j^2}{2}
 +\eta_j\ln\eta_j+\left(1-\eta_j\right)\ln\left(1-\eta_j\right)\Bigg],
\label{eq:2_SI}
\end{align}
where $\kappa^{-1}=\sqrt{\varepsilon/\left(2\beta e^2I\right)}$ is the 
Debye length with $I$ being the ionic strength. This is the expression 
for the grand potential we provide in the main text.  


\begin{centering}
\subsection*{2. Minimization of the grand potential}
\end{centering}

The integral in the first line of Eq.~\eqref{eq:2_SI} can be decomposed 
in the following way:
\begin{align}
\beta\Omega\left[\sigma^{*}\right]=&-\frac{\varepsilon}{\beta e^2}\int\limits_{-\infty}^{0}dz
 \left[\kappa^2\left(\cosh\left(\phi(z)\right)-1\right)+\frac{1}{2}\left(\phi'(z)\right)^2\right]\nonumber\\
&-\frac{\varepsilon}{\beta e^2}\int\limits_{0}^{L}dz
 \left[\kappa^2\left(\cosh\left(\phi(z)\right)-1\right)+\frac{1}{2}\left(\phi'(z)\right)^2\right]\nonumber\\
&-\frac{\varepsilon}{\beta e^2}\int\limits_{L}^{\infty}dz
 \left[\kappa^2\left(\cosh\left(\phi(z)\right)-1\right)+\frac{1}{2}\left(\phi'(z)\right)^2\right]\nonumber\\
&+\frac{1}{a^2}\sum\limits_{j=1}^{N}\Bigg[\sigma_{j}^*\phi_j
 -\alpha\eta_j-\frac{\chi \eta_j^2}{2}
 +\eta_j\ln\eta_j+\left(1-\eta_j\right)\ln\left(1-\eta_j\right)\Bigg].
\label{eq:3_SI}
\end{align}
First, we consider the first integral (from $-\infty$ to $0$) in 
Eq.~\eqref{eq:3_SI}. One can rewrite it as 
\begin{align}
I_1&=-\frac{\varepsilon}{\beta e^2}\int\limits_{-\infty}^{0}dz
 \left[\kappa^2\left(\cosh\left(\phi(z)\right)-1\right)+\frac{1}{2}\left(\phi'(z)\right)^2\right]\nonumber\\
 &=-\int\limits_{-\infty}^{0}dz
 \left[2I\left(\cosh\left(\phi(z)\right)-1\right)+\frac{\varepsilon}{2\beta e^2}\left(\phi'(z)\right)^2\right].
\label{eq:4_SI}
\end{align}
Multiplying both sides of the PB equation by 
$\phi'$ one obtains
\begin{align}
 \phi'\phi''=\kappa^2\sinh\left(\phi\right)\phi',
\label{eq:5_SI}
\end{align}
which can be rewritten as
\begin{align}
 \frac{1}{2}\left(\left(\phi'\right)^2\right)'=\kappa^2\left(\cosh\left(\phi\right)\right)'.
\label{eq:6_SI}
\end{align}
Integrating Eq.~(\ref{eq:6_SI}) with respect to $z$ and using $\phi\left(-\infty\right)=
\phi'\left(-\infty\right)=0$ gives
\begin{align}
 \frac{1}{2}\left(\phi'\right)^2=\kappa^2\left(\cosh\left(\phi\right)-1\right),
\label{eq:7_SI}
\end{align}
which leads to
\begin{align}
 2I\left(\cosh\left(\phi\right)-1\right)=\frac{\varepsilon}{2\beta e^2}\left(\phi'\right)^2.
\label{eq:8_SI}
\end{align}
Using Eq.~\eqref{eq:8_SI}, Eq.~\eqref{eq:4_SI} simplifies to
\begin{align}
I_1=-\int\limits_{-\infty}^{0}dz\frac{\varepsilon}{\beta e^2}\left(\phi'(z)\right)^2.
\label{eq:9_SI}
\end{align}
If we consider only the semi-infinite geometry from $-\infty$ to $0$ 
having a charged surface at $z=0$, the PB equation for this setup is 
analytically solvable and its solution is well known \cite{Hun89, Rus89}:
\begin{align}
 \phi(z)=4\operatorname{artanh}\left(\gamma\exp\left(\kappa z\right)\right);
 ~~~~~\gamma=\tanh\left(\frac{\phi(0)}{4}\right).
\label{eq:10_SI}
\end{align}
Taking the derivative with respect to $z$, one obtains
\begin{align}
 \phi'(z)=4\kappa\gamma\frac{\exp\left(\kappa z\right)}{1-\gamma^2\exp\left(2\kappa z\right)}.
\label{eq:11_SI}
\end{align}
Therefore,
\begin{align}
 \int\limits_{-\infty}^0dz\left(\phi'(z)\right)^2&=16\kappa^2\gamma^2\int\limits_{-\infty}^0dz
         \frac{\exp\left(2\kappa z\right)}{\left(1-\gamma^2\exp\left(2\kappa z\right)\right)^2}\notag\\
        &=8\kappa\int\limits_{-\infty}^0dz
         \frac{2\kappa\gamma^2\exp\left(2\kappa z\right)}{\left(1-\gamma^2\exp\left(2\kappa z\right)\right)^2}\notag\\
        &=8\kappa\int\limits_{-\infty}^0dz
         \left(\frac{d}{dz}\frac{1}{1-\gamma^2\exp\left(2\kappa z\right)}\right)\notag\\
        &=8\kappa\left|\frac{1}{1-\gamma^2\exp\left(2\kappa z\right)}\right|_{z=-\infty}^{0}\notag\\
        &=8\kappa\left(\frac{1}{1-\gamma^2}-1\right)\notag\\
        &=\frac{8\kappa\gamma^2}{1-\gamma^2}.
\label{eq:12_SI}
\end{align}
Using this, Eq.~\eqref{eq:9_SI} becomes
\begin{align}
I_1&=-\frac{\varepsilon}{\beta e^2}\frac{8\kappa\gamma^2}{1-\gamma^2}\notag\\
   &=-\frac{8\kappa\varepsilon}{\beta e^2}
   \frac{\tanh\left(\frac{\phi(0)}{4}\right)^2}{1-\tanh\left(\frac{\phi(0)}{4}\right)^2}\notag\\
   &=-\frac{8\kappa\varepsilon}{\beta e^2}\sinh\left(\frac{\phi(0)}{4}\right)^2.
\label{eq:13_SI}
\end{align}
Similarly, it can be shown that the integral from $L$ to $\infty$ in 
Eq.~\eqref{eq:3_SI} boils down to
\begin{align}
I_2=-\frac{8\kappa\varepsilon}{\beta e^2}\sinh\left(\frac{\phi(L)}{4}\right)^2.
\label{eq:14_SI}
\end{align}
Using Eqs.~\eqref{eq:13_SI} and \eqref{eq:14_SI}, in Eq.~\eqref{eq:3_SI}, 
one arrives at the expression
\begin{align}
\beta\Omega\left[\sigma^{*}\right]=&-\frac{8\kappa\varepsilon}{\beta e^2}\left[\sinh\left(\frac{\phi(0)}{4}\right)^2
+\sinh\left(\frac{\phi(L)}{4}\right)^2\right]\nonumber\\
&-\frac{\varepsilon}{\beta e^2}\int\limits_{0}^{L}dz
 \left[\kappa^2\left(\cosh\left(\phi(z)\right)-1\right)+\frac{1}{2}\left(\phi'(z)\right)^2\right]\nonumber\\
&+\frac{1}{a^2}\sum\limits_{j=1}^{N}\Bigg[\sigma_{j}^*\phi_j
 -\alpha\eta_j-\frac{\chi \eta_j^2}{2}
 +\eta_j\ln\eta_j+\left(1-\eta_j\right)\ln\left(1-\eta_j\right)\Bigg].
\label{eq:15_SI}
\end{align}
This is the expression we minimize to obtain the equilibrium profiles 
for $\sigma^*=(\sigma_{\!1}^*,\dots,\sigma_{\!N}^*)$. The minimization 
is done by two nested loops: the inner one calculates $\phi(z)$ for 
fixed $\sigma^*$ profiles and the outer one performs a steepest descent 
step to minimize $\Omega$. The iteration is done until convergence is 
reached.

\begin{figure}[!t]
\begin{center}
\includegraphics[width=1\columnwidth]{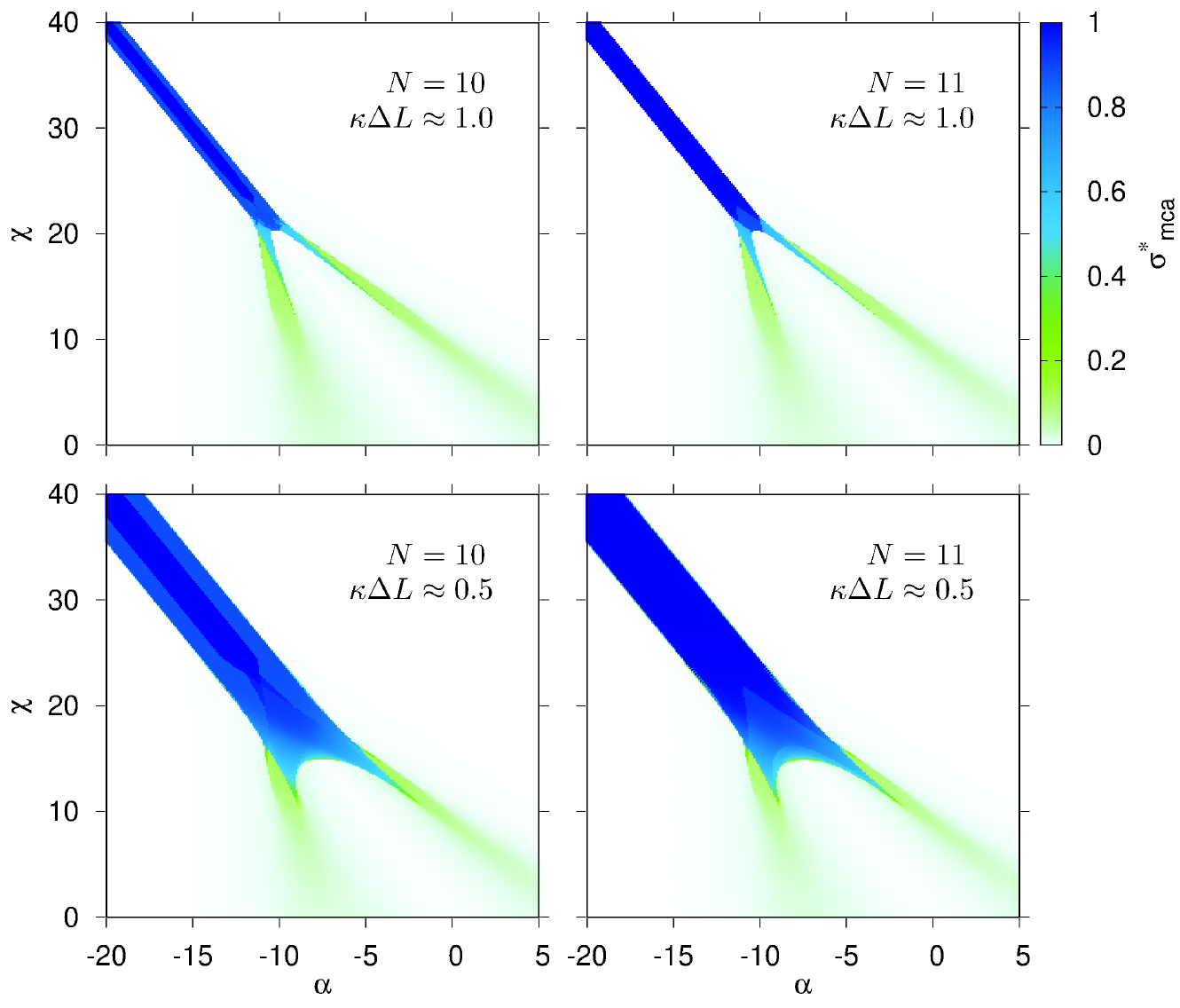}
\caption{Variation of the quantity $\sigma_{\text{mca}}^*$ for stacks of 
         $N=10$ (left panel) and $N=11$ (right panel) charge-regulated 
         interacting surfaces with the interaction parameters $\alpha$ 
         and $\chi$ and with varying dimensionless separation 
         $\kappa\Delta L$ between two consecutive surfaces. As one 
         can see, with decreasing $\kappa\Delta L$ the dark blue 
         region with the highest charge-asymmetry spreads.}
\label{Fig1_SI}
\end{center}
\end{figure}


\begin{centering}
\subsection*{3. Variation of the stack width}
\end{centering}

In the main text, we have shown results for the case of 
separation $\kappa\Delta L\approx1$ between two consecutive 
surfaces in the stack. When this separation changes, the 
electrostatic interaction between the surfaces, which is at 
the origin of the observed charge asymmetry \cite{Maj18}, also 
changes. As a result, for given $\alpha$ and $\chi$, the 
charge distribution inside the stack changes with varying 
$\kappa\Delta L$. This can be inferred from Fig.~\ref{Fig1_SI} 
where we have shown the quantity $\sigma^*_\text{mca}$ for two 
different $\kappa\Delta L$ values. With decreasing separation 
between the surfaces, their electrostatic attraction becomes 
stronger and it dominates over other interactions present 
in the system for a wider range of $\alpha$ and $\chi$ parameters. 
Of course, the opposite happens when increasing the inter-surface 
separation: the electrostatic interaction becomes weaker and the 
colored region of asymmetric charge distributions becomes narrower.

As discussed in the main text, for an even number of surfaces 
inside the stack, a reorganization of the charge distribution 
takes place across the line $\chi=-2\alpha$ and consequently, 
$\sigma_{\text{mca}}^*$ changes color (see Fig.~\ref{Fig4} of 
the main text). With increasing number $N$ of surfaces inside 
the stack, this feature remains as it has nothing to do with 
total number of surfaces inside the stack as long as $N$ is 
even. However, the contrast in $\sigma_{\text{mca}}^*$ values 
becomes lower as the denominator in Eq.~(2) of the main text 
increases with increasing $N$. As a result, the color contrast 
within the blue region in the left panel of Fig.~\ref{Fig1_SI} 
is lower compared to that in Fig.~\ref{Fig4} of the main text.


\begin{centering}
\subsection*{4. Disjoining pressure in the small stack width limit}
\end{centering}
As mentioned in the main text, all the three curves in 
Fig.~\ref{Fig5} increase linearly for very short stack widths 
$\kappa L\lesssim 4$. At such short separations, electrostatics 
is almost unscreened and each oppositely and highly charged 
pair of surfaces (with charge densities 
$|\sigma|\approx 0.5\,e/\mathrm{nm}^2$) inside the stack acts 
like a capacitor. So the total energy of the system is, besides 
the contribution of the outer parts, a sum of these capacitor 
energies. Each surface contributes to its both sides; therefore, 
half of their charges contribute to each capacitor. As a result, 
the $L$-dependent part of the grand potential per unit 
cross-sectional area equals
\begin{align}
\Omega=\frac{\sigma^2\Delta L(N-1)}{8\varepsilon_0\varepsilon_r}=\frac{\sigma^2L}{8\varepsilon_0\varepsilon_r}
\label{eq:16_SI}
\end{align}
The factor of `$8$' in the denominator results from the fact 
that half of the total charge on each surface contributes to 
a capacitor. The corresponding disjoining pressure 
$-\frac{\partial(\Omega)}{\partial(L)}$ then equals to 
$-\frac{\sigma^2}{8\varepsilon_0\varepsilon_r}=-0.273\,k_BT/\mathrm{nm}^3$, 
which, when divided by $\kappa\approx 0.1027\,\mathrm{nm}^{-1}$, 
matches perfectly with the observed slope $(2.66\,k_BT/\mathrm{nm}^2)$ 
of the linear parts of the curves in Fig.~\ref{Fig5}. 



\begin{thebibliography}{00}
\bibitem{finite} The underpinnings and limitations of this approximation are discussed in Ref.~\cite{Mar19}.
\end{thebibliography}
\end{document}